
\input phyzzx.tex
\def\fa {{\delta f_0}}
\def\fb {{\delta f_1}}
\def\fc {{\delta f_2}}
\def\fd {{\delta f_3}}
\def\fx {{\delta \Phi}}
\def\tfa {{\delta \tilde f_0}}

\def\tfc {{\delta \tilde f_2}}

\def\tfx {{\delta \tilde \Phi}}
\def\mr {{\left (1-{M\over r}\right )}}
\def\qr {{\left (1-{Q^2\over Mr}\right )}}
\def\qm {{\left (1-{Q^2\over M^2}\right )}}
\def\htdz {{\delta H_{320}+{Q\over r^2}\chi_3}}
\def\derp {{\partial_+}}
\def\derm {{\partial_-}}
\def\dr {{\partial_{r^*}}}
\def\ddr {{\partial^2_{r^*}}}
\pubnum{ROM2F-92/18}
\titlepage
\title{ On the Stability of a Stringy Black Hole}
\author{ A. Carlini}
\address{SISSA, Strada Costiera 11, 34014 Trieste,
Italy and INFN, sezione di Genova, Via Dodecaneso 33, 16146 Genova}
\author{ F. Fucito \foot{email addresses: vaxtov::fucito and
fucito@roma2.infn.it}}
\address{Dipartimento di Fisica, Universit\`a
di Roma II ``Tor Vergata" and INFN, sezione di Roma  ``Tor Vergata",
Via Carnevale, 00173 Roma, Italy}
\andauthor{ M. Martellini \foot{Permanent address: Dipartimento di Fisica,
Universit\`a di Milano, 20133 Milano, Italy and INFN, Sezione di Pavia,
Via Bassi 6, 27100 Pavia, Italy}}
\address{INFN, sezione di Roma ``La Sapienza", P.le A. Moro 2, 00185 Rome,
Italy}
\unnumberedchapters
\abstract
We study the stability under perturbations of a charged four dimensional
stringy black hole arising from gauging a previously studied WZW model.
We find that the black hole is stable only in the extremal case $Q=M$.
\endpage
\pagenumber=1

If string theory is to describe a quantum theory of gravity, it is certainly
important to investigate what happens to it around singular backgrounds
and how
such backgrounds are generated. Addressing the first issue leads to the
conclusion that in certain cases the propagation of strings through backgrounds
which are singular in the sense of classical gravity, is followed by the
excitation
of an infinite number of modes of the string itself\rlap.\Ref\sinback{G.
Horowitz and A.Steif\journal Phys.Rev.Lett.&64(90)260;
Phys.Rev.Lett.{\bf 65}\break
(1990)\b1518; Phys.Lett.{\bf B258}(1991)91;\nextline
H.de Vega and Sanchez\journal Phys.Rev.Lett.&65(90)1517.}
For what the second
problem is concerned, black hole type singularities were first obtained by
satisfying the equations of motion arising from the four dimensional low-energy
Lagrangian, obtained from string theory, which
describe the coupling of gravitational, dilatonic, Maxwell and antisymmetric
fields\rlap.\Ref\gibbon{G.W.Gibbons and K.Maeda\journal Nucl.Phys.&B298(88)741;
\nextline
D.Garfinkle, G.Horowitz and A.Strominger\journal Phys.Rev.&43D(91)3140;
\nextline C.G.Callan, R.C.Myers and M.J.Perry\journal Nucl.Phys.&B311(88/89)
673.}

Recently a new way to generate 2-D black hole backgrounds has been
devised\Ref\witten{E.Witten\journal Phys.Rev.&44D(91)314; \nextline
R.Dijkgraaf, H.Verlinde and E.Verlinde\journal Nucl.Phys.&B371(92)269.}
gauging a WZW model built on a coset manifold.

The black holes obtained in Ref.\gibbon~have been throughly analized in a
recent series of papers\rlap.\Ref\preskill{J.Preskill, P.Schwarz,
A.Shapere, S.Trivedi and F.Wilczek\journal Mod.Phys.Lett.&A6(91)2353;
\nextline A.Shapere, S.Trivedi and F.Wilczek, {\it Dual Dilaton Dyons},
IASSNS-HEP-91/33, June 1991;  \nextline
S.B.Giddings and A.Strominger, {\it Dynamics of Extremal Black Holes},
UCSB-TH-92-01, February 1992.}\Ref\wil{C.Holzhey and F.Wilczek, {\it Black
Holes as Elementary Particles}, IASSNS-HEP-91/71, December 1991.}
Quite surprisingly, the black holes obtained
in this fashion exhibit different thermodynamical properties from those
obtained from
general relativity. In some instances, for example, their scattering
behaviour is more similar to that of an elementary particle than a
thermal object.

\REF\gilbert{G.Gilbert, {\it On the Perturbations of String Theoretic Black
Holes}, UMDEPP 92-035, August 1991; {\it The Instability of String
Theoretic Black Holes}, UMDEPP 92-110, November 1991.}
\REF\raiten{E.Raiten, {\it Perturbations of a Stringy Black Hole},
FERMI-PUB-91/338-T.}
The purpose of this letter is to analyze the behaviour of a four
dimensional black hole obtained along the lines of Ref.\witten: we study
how it behaves under geometrical perturbations and we will briefly describe
its thermodynamical
properties which turn out (in the extremal case) to be different both
from those of black holes obtained from classical gravity and those of
Ref.\preskill. The study of perturbations of stringy black holes
in two and four dimensions has been
carried out in Ref.\gilbert, \wil, \raiten. Our results are
different from those of Ref.\raiten: the black hole under study is stable
under perturbations of the metric
only in the extremal case, thus supporting the conjecture that extremal
black holes might be stable ``quantum" ground state for the underlying theory.

\REF\horowitz{J.Horne and G.Horowitz, {\it Exact Black String Solutions in
Three Dimensions}, UCSBTH-91-39, July 1991.}
We now turn to the derivation of a four dimensional black hole
following Ref.\raiten, \horowitz.
The starting point of our analysis is the WZW action:
$$
L(g)={k Tr\over 4\pi}\left [\int_{\Sigma}d^2\sigma \left (g^{-1}\derp
gg^{-1}\derm g \right )-\int_B{d^3y \over 3}\left (g^{-1}dg\wedge g^{-1}dg
\wedge g^{-1}dg \right )\right ],
\eqn\uno
$$
where the integrals are over the three manifold $B$ and its boundary
$\Sigma$.
We now gauge a one dimensional subgroup $H$ of the symmetry group, with
action $g\rightarrow hgh^{-1}$, and introduce the gauge field $A_i$, whose
gauge transformations are:
$$
\cases{\delta a=2\epsilon a, &\cr
\delta b=-2\epsilon b, &\cr
\delta u=\delta v=0, &\cr
\delta x_i=2\epsilon c_i, &\cr
\delta A_i=-\partial_i \epsilon. &\cr}
\eqn\due
$$
Raiten proposal follows by adding two free bosons $x_1$ and $x_2$ to the
2-d black hole of Ref. \witten, that is by letting $G=SL(2, R)\times R \times
R$, and by modding out the translations in both $x_1$ and $x_2$.
Parametrizing $SL(2, R)$ as:
$$
g=\pmatrix{a&u\cr
-v&b\cr},
\eqn\tre
$$
the gauged WZW action which is invariant under \due~becomes:
$$
\eqalign{L(g,A)=L(g)&+{k\over 2\pi}\int d^2\sigma A_+\left (b\derm a
-a\derm b -u\derm v +v\derm u +{4c_i\over k}\derm x_i\right )\cr
&+{k\over 2\pi}\int d^2\sigma A_-\left (b\derp a -a\derp b +u\derp v -v\derp
u +{4c_i\over k}\derp x_i\right )\cr
&+{2k\over \pi}\int d^2\sigma A_+A_-\left ( 1+{2c^2\over k}-uv\right ),\cr}
\eqn\quattro
$$
where a sum over $i=1, 2$ is assumed, and $c_1, c_2$ are constants such
that:
$$
c^2=c_1^2+c_2^2.
\eqn\pippa
$$
We then fix the gauge by setting $a=\pm b$, depending on the sign of
$1-uv$, and we choose to work in the ansatz:
$$
c_1=c_2={\sqrt {k}\over 2}\left ({M^2 \over Q^2}-1\right )^{-1/2}.
\eqn\sei
$$
By making the transformation of variables:
$$
\cases{u=e^{\sqrt {{2\over k}}\qm ^{1/2}t}\left ({r\over M}
-1\right )^{1/2}\qm ^{-1/2}&, \cr
v=-e^{-\sqrt {{2\over k}}\qm ^{1/2}t}\left ({r\over M}
-1\right )^{1/2}\qm ^{-1/2}&, \cr}
\eqn\sette
$$
and integrating out the gauge fields, the WZW action finally turns out:
$$
L={1\over \pi}\int ~d^2\sigma ~\left [g_{\mu \nu}\derm x^{\mu}\derp x^{\nu}~+
{1\over 2}B_{\mu \nu}(\derm x^{\mu}\derp x^{\nu}-\derp x^{\mu}\derm x^{\nu})
\right ],
\eqn\otto
$$
where $g_{\mu \nu}$ is the 4-d metric, $B_{\mu \nu}$ an antisymmetric
tensor field, and $x^{\mu}=(t, r, x_1, x_2)$.
It is then almost straightforward to show that requiring the fields to be an
extremum of the low energy effective action from string theory:
$$
S=\int ~d^4x\sqrt{-g}~e^{\Phi}\left [R+(\nabla \Phi)^2-{H^2\over 12}+{8\over k}
\right ],
\eqn\nove
$$
and redefining $x_1, x_2$ coordinates as:
$$
\cases{x_1={1\over \sqrt{2}}(x+y),\cr
x_2={1\over \sqrt{2}}(x-y),\cr}
\eqn\dieci
$$
the final form of the fields is:
$$
\eqalign{&ds^2=-\mr dt^2+\qr dx^2+dy^2+{k~dr^2\over 8(r-M)(r-Q^2/M)}, \cr}
\eqn\undicia
$$
$$
\eqalign{&H_{rtx}={Q\over r^2}, \cr}
\eqn\undicib
$$
$$
\eqalign{&\Phi =\ln (r) +{1\over 2}\ln \left ({k\over 2}\right ), \cr}
\eqn\undicic
$$
where $H\dot = d~B$.
The field equations coming from this effective action are\rlap:
\foot{Our equation (16) contains a $H^2_{\mu\nu}$ term which is not
present in (4.3) of Ref.\raiten~and which will make the final equations
look quite different.}
$$
\eqalign{&\nabla_\lambda (e^{\Phi}H^{\lambda \mu \nu})=0, \cr}
\eqn\dodicia
$$
$$
\eqalign{&-{H^2\over 6}+\nabla^2\Phi +(\nabla \Phi)^2-{8\over k}=0, \cr}
\eqn\dodicib
$$
$$\eqalign{&
R_{\mu \nu}=\nabla_{\mu}\nabla_{\nu}\Phi+{1\over 2}g_{\mu \nu}\left (
\nabla^2\Phi+(\nabla \Phi)^2-{8\over k}-{H^2\over 6}\right )+{H_{\mu
\nu}^2\over 4}~\dot =~ T_{\mu \nu}, \cr}
\eqn\dodicic
$$
$$
\eqalign{&
dH=H_{\mu \nu \lambda ,\rho}-H_{\nu \lambda \rho ,\mu}+H_{\lambda \rho \mu,
\nu}-H_{\rho \mu \nu ,\lambda}=0. \cr}
\eqn\dodicid
$$

Let us now summarize the global structure of the above metric:
\item
\bullet
$Q>M$. Contrary to the charged black holes of general relativity, in this
regime our black hole exhibits neither a horizon nor a curvature
singularity (naked singularity).
\item \bullet
$Q<M$. The solution has a curvature singularity and two Killing horizons: an
outer horizon at $r=r_+=M$ and an inner horizon at $r=r_-={Q^2\over M}$.
Opposed to the general relativity black hole the generator of time
translations, remains space-like also for $r<r_-$. As a consequence the
manifold is time-like and light-like geodesically complete.
\item \bullet
$Q=M$. This is the extremal case in which $r_+=r_-$. With respect to the
general relativity solution, we notice that the metric is boosted
along the $x$ direction.

We now discuss the thermodynamics.
The temperature and entropy of our black hole are:
$$
T={1\over \pi M}\sqrt{{M^2-Q^2\over 2k}},
\eqn\tredicibis
$$
$$
S={A\over 4}\biggr \vert_{r_+}={\pi^2\over 2}\left (1-{r_-\over r_+}\right )
^{1/2},
\eqn\tredici
$$
where the coordinates $x, y$ are now periodic: $x\in [0, 2\pi ]$,
$y\in [0, \pi ]$ and $A$ is the horizon area.
We here remark the difference with the other stringy black hole of
Ref.\preskill~which has a temperature $T={1 \over 8\pi M}$, independent of
the charge.

In the extremal case the black hole under study has zero entropy and
temperature while the classical gravity (string) solution has zero (${1
\over 8\pi M}$) temperature and finite (zero) entropy\rlap.\foot{The
authors of Ref.\gibbon~consider also a model with a parameter $a$ which
interpolates between the classical gravity case ($a=0$) and the string case
($a=1$). The thermodynamical properties of the black hole under study are
thus equivalent to the case $0<a<1$.}

Let us now investigate the range of validity of the thermal description of
the black hole defined by \undicia. According to Ref.\preskill:
$$
{\partial T\over \partial M}\biggr \vert_{Q}={Q^2\over \pi
\sqrt{2k}M^2(M^2-Q^2)^{1/2}}\gg 1,
\eqn\quattordici
$$
for the thermal description to break down. This is the case for the
extremal hole where $Q\longrightarrow M$. This is true independently of the
value of the mass similarly to what happens to the black hole of
Ref.\preskill~but in contrast with the extreme Reissner-Nordstr\"om
solution.\REF\birrel{N.D.Birrel and P.C.W.Davis, {\it Quantum Fields in
Curved Space}, (Cambridge University Press ,Cambridge, 1982).}
Following Ref.\birrel~we now discuss the domain of validity of the
semi-classical approximation. This approximation breaks down when:
$$
{1 \over M} {\partial M \over \partial t} \simeq T.
\eqn\birreldavis
$$
Using the Stefan-Boltzmann radiation law, this implies $T M \simeq A  T^4$.
In the extremal limit $T\mapsto 0$ and the previous formula is satisfied
independently of the value of the mass.

Let us now study the perturbations of the metric field. We rewrite the
metric as:
$$
g_{\mu \nu}=\pmatrix{-e^{2f_0}&0&0&0\cr
0&e^{2f_1}&0&0\cr
0&0&e^{2f_2}&0\cr
0&0&0&e^{2f_3}\cr },
\eqn\quindici
$$
and let us take as an {\it ansatz} for the axial and polar perturbations a
sufficiently general definition consistent with time-dependence and axial
simmetry:
$$
\delta g_{\mu \nu}=\delta g^A_{\mu \nu}+\delta g^P_{\mu \nu}=\pmatrix{-2\fa
e^{2f_0}&-\chi_0 e^{2f_1}&0&0\cr
-\chi_0 e^{2f_1}&2\fb e^{2f_1}&-\chi_2 e^{2f_1}&-\chi_3 e^{2f_1}\cr
0&-\chi_2 e^{2f_1}&2\fc e^{2f_2}&0\cr
0&-\chi_3 e^{2f_1}&0&2\fd e^{2f_3}\cr},
\eqn\sedicibis
$$
with $x^{\mu}=(t, x, y, r)$. In view of our choice, the first order
perturbations $\delta f_{\mu}, \chi_{\mu}$ are $x$ independent.

\REF\chandra{S.Chandrasekhar, {\it The Mathematical Theory of Black Holes},
(Clarendon Press, Oxford, 1983).}
We now have to compute the first order variation of the equation of motion
\dodicia-\dodicid. Following Ref.\chandra~we compute the variation of the
energy-momentum tensor in the tetrad formalism. We thus rewrite the metric as:
$g_{\mu \nu}=e^{(a)}_{\mu}e^{(b)}_{\nu}\eta_{(a)(b)}$.
The variations of the tetrads are:
$$
\eqalign{&\delta e^{\mu}_{(0)}=(-\fa e^{-f_0},~\chi_0 e^{-f_0},~0,~0),\cr
&\delta e^{\mu}_{(1)}=(0,~~-\fb e^{-f_1},~~~~0,~~~~0),\cr
&\delta e^{\mu}_{(2)}=(0,~\chi_2 e^{-f_2},~-\fc e^{-f_2},~0),\cr
&\delta e^{\mu}_{(3)}=(0,~\chi_3 e^{-f_3},~0,~-\fd e^{-f_3}).\cr}
\eqn\sedici
$$
The variation of the components of the energy-momentum tensor are:
$$
\eqalign{&\delta T_{(1)(2)}=0, \cr}
\eqn\disettea
$$
$$
\eqalign{
&\delta T_{(1)(3)}={e^{-2f_2}\over 2r}\left [ e^{-2f_0-f_1}{Q\over r}\left (
\htdz \right )+e^{f_1}\chi_{23}\right ], \cr}
\eqn\disetteb
$$
$$
\eqalign{
&\delta T_{(0)(0)}=e^{-2f_0-2f_1-2f_2}\left [{Q\over r^2}\delta H_{012}-{Q^2
\over r^4}(\fa +\fb +\fc)\right ]+ \cr
&-e^{-2f_2}\biggl (f_{0,2}\fx_{,2}+
{\fa_{,2}\over r}\biggr )-\Delta, \cr}
\eqn\disettec
$$
$$
\eqalign{
&\delta T_{(1)(1)}={Q^2\over r^4}e^{-2f_0-2f_1-2f_2}(\fa +\fb +\fc)+e^{-2f_2}
\biggl (f_{1,2}\fx_{,2} + \cr
&+{\fb_{,2}\over r}\biggr )-{8Q\over k}\delta H_{012}+\Delta, \cr}
\eqn\disetted
$$
$$
\eqalign{
&\delta T_{(2)(2)}={Q^2\over r^4}e^{-2f_0-2f_1-2f_2}(\fa +\fb +\fc))+{e^{-2f_2}
\over r^2} [2\fc (1+rf_{2,2})+ \cr
&-\fc_{,2} r +r^2(\fx_{,2,2}-f_{2,2}\fx_{,2})]-{8Q\over k}\delta H_{012}
+\Delta, \cr}
\eqn\disettee
$$
$$
\eqalign{
&\delta T_{(3)(3)}={e^{-2f_2-2f_3}\over r}\fd_{,2}+e^{-2f_3}\fx_{,3,3}+\Delta,
 \cr}
\eqn\disettef
$$
$$
\eqalign{
&\delta T_{(0)(2)}=-{e^{-f_0-f_2}\over r}(\fc_{,0}+rf_{0,2}\fx_{,0}), \cr}
\eqn\disetteg
$$
$$
\eqalign{
&\delta T_{(0)(3)}={Q\over 2r^2}e^{-f_0-2f_1-2f_2}\delta H_{123}, \cr}
\eqn\disetteh
$$
$$
\eqalign{
&\delta T_{(2)(3)}=-{Q\over 2r^2}e^{-2f_0-2f_1-f_2}\delta H_{013}-{e^{-f_2}
\over r}\fc_{,3}, \cr}
\eqn\disettei
$$
where:
$$\eqalign{
&\Delta\dot=
-{1\over 2}e^{-2f_0}\fx_{,00}+{4r^2\over k}e^{2f_0+2f_1} \fx_{,22}+{1\over 2}
\fx_{,33}+{4r\over k}\biggl (3-{2M\over r}+ \cr
&-{2Q^2\over Mr}+{Q^2\over r^2}\biggr )\fx_{,2}+
{4r\over k}e^{2f_0+2f_1}(\fa +\fb -\fc +\fd )_{,2}-{8\over k}\fc + \cr
&-{8Q^2\over kr^2}(\fa +\fb)+{8Q\over k}\delta H_{012}. \cr}
\eqn\pippo
$$
After defining $\chi_{\alpha \beta}\dot =\chi_{\alpha, \beta}-
\chi_{\beta, \alpha}$, we are now ready to write all the perturbation
equations at first order:
$$
\eqalign{&\chi_{23,3}-e^{-2f_0}\chi_{20,0}=0, \cr}
\eqn\diciannovea
$$
$$
\eqalign{
&e^{-2f_1-f_0+f_2}\left [ \left ( e^{3f_1+f_0-f_2}\chi_{23}\right )_{,2}+
\left ( e^{3f_1-f_0+f_2}\chi_{30}\right )_{,0}\right ]= \cr
&=-{1\over r}\left [
e^{-2f_0-f_1}{Q\over r}\left (\htdz \right )+ e^{f_1}\chi_{23}\right ], \cr}
\eqn\diciannoveb
$$
$$
\eqalign{
&e^{-2f_0}\fb_{,00}-{8r^2\over k}e^{2f_0+2f_1}\fb_{,22}-\fb_{,33}
+{4Q^2\over Mk}e^{2f_0}(\fc -\fd
-\fa )_{,2}+ \cr
&-{8r\over k}\left (1+{Q^2\over 2Mr}-{3Q^2\over  2r^2}\right )
\fb_{,2}-{8Q^2\over kMr}\left (1-{2M\over r} -{Q^2 \over Mr}+
2{Q^2 \over r^2}\right ) e^{-2f_1}\fc= \cr
 {8Q\over k}\biggl [{Q\over r^2}(\fa+
&+\fb +\fc)-\delta H_{012}\biggr ]
+{4Q^2\over Mk}e^{2f_0}\fx_{,2}+{8r\over k}e^{2f_0+2f_1} \fb_{,2}, \cr}
\eqn\diciannovec
$$
$$
\eqalign{
&e^{-2f_0}\fc_{,00}-{8r^2\over k}e^{2f_0+2f_1}(\fa +\fb +\fd )_{,22}
-\fc_{,33}-{8r\over k}\biggl (1-{Q^2\over 2Mr}+ \cr
&+{M\over 2r}-{Q^2\over r^2}\biggr )\fa_{,2}
+{4M\over k}\left (1+{Q^2\over M^2}-{2Q^2\over Mr}\right )\fc_{,2}
-{4r\over k}\biggl (2-{M\over r}+ \cr
&-{Q^2\over Mr}\biggr )\fd_{,2}-{8r\over k}
\biggl (1-{M\over 2r}+{Q^2\over 2Mr}
-{Q^2\over r^2}\biggr )\fb_{,2}+{8M^2\over kr^2}\biggl (1+{Q^4\over M^4}+ \cr
&+{5Q^2\over M^2}-{r\over M}+{3Q^4\over M^2r^2}-{Q^2r\over M^3}
-{4Q^4\over M^3r}-{4Q^2\over Mr}\biggr )e^{-2f_0-2f_1}\fc
= \cr
&={8Q\over k}\biggl [{Q\over r^2}(\fa+\fb)+{1\over Q}\left (
{3Q^2\over r^2}-{M\over r}-{Q^2\over Mr}\right )\fc -\delta H_{012}\biggr ]+
\cr
&+ {8r\over k}e^{2f_0+2f_1}(r\fx_{,22}-\fc_{,2})
+{4r\over k}\left (2-{M\over r}-{Q^2\over Mr}\right )\fx_{,2}, \cr}
\eqn\diciannoved
$$
$$
\eqalign{
&-e^{-2f_0}(\fb +\fc +\fd )_{,00}+{8r^2\over k}e^{2f_0+2f_1}\fa_{,22}
+{8r\over k}\biggl (1+{M\over 2r}+ \cr
&-{3Q^2\over 2r^2}\biggr )\fa_{,2}
+\fa_{,33}+{4M\over k}e^{2f_1}(\fb -\fc +\fd )_{,2}
+{8M\over kr}\biggl (1- {M\over r}+ \cr
&-{2Q^2\over Mr}+{2Q^2\over
r^2}\biggr )e^{-2f_0}\fc = {8\over k}\biggl [Q \biggl [
\delta H_{012}-{Q\over r^2}(\fa
+\fb+\fc)\biggr ]+ \cr
&-re^{2f_0+2f_1} \biggl [{M\over 2r}
{\fx_{,2}\over e^{2f_0}}+\fa_{,2}\biggr ]\biggr ], \cr}
\eqn\diciannovee
$$
$$
\eqalign{
&e^{-2f_0}\fd_{,00}-{8r^2\over k}e^{2f_0+2f_1}\fd_{,22}-(\fa +\fb +\fc )_{,33}
= \cr
&={8r\over k}e^{2f_0+2f_1}\fd_{,2} +\fx_{,33}, \cr}
\eqn\diciannovef
$$
$$
\eqalign{
&-re^{f_1}(\fb +\fd )_{,02}+{1\over 2}{Q^2\over Mr}e^{-f_1}\fc_{,0}
+{M\over 2r}e^{-2f_0}\biggl [e^{f_1}\fd_{,0}+ \cr
&+\qm e^{-f_1}\fb_{,0}\biggr ]=-e^{f_1}\biggl [\fc_{,0}+{M\over 2r}e^{-2f_0}
\fx_{,0}\biggr ], \cr}
\eqn\diciannoveg
$$
$$
\eqalign{
&-e^{-f_0}(\fb +\fc )_{,03}={4Q\over k}e^{f_0}\delta H_{123}, \cr}
\eqn\diciannoveh
$$
$$
\eqalign{
&-e^{f_1}\left (r e^{f_0}(\fa +\fb )_{,23}+{1\over 2}{M\over r}e^{-f_0}
\fa_{,3}\right )+{1\over 2}e^{-f_1}\biggl [\biggl (1-{2Q^2\over Mr}+ \cr
&+{Q^2\over M^2}\biggr ){M\over r}e^{-f_0}\fc_{,3}
-{Q^2\over Mr}e^{f_0}\fb_{,3}\biggr ]= -e^{f_0+f_1}\fc_{,3}
-{Q\over 2r}e^{-f_0-f_1}\delta H_{013}, \cr}
\eqn\diciannovei
$$
$$
\eqalign{
&{k\over 8}e^{-2f_0-2f_1}\delta H_{013,3}+\biggl [r^2
\delta H_{012}+Q(\fd +\fx -\fa -\fb -\fc)\biggr ]_{,2}=0, \cr}
\eqn\diciannovel
$$
$$
\eqalign{
&e^{2f_0}\delta H_{123,3}-\left [\delta H_{012}+{Q\over
r^2}(\fd +\fx -\fa -\fb -\fc)\right ]_{,0}=0, \cr}
\eqn\diciannovem
$$
$$
\eqalign{
&\delta H_{013,0}+{8r\over k}e^{2f_0+2f_1}\left [re^{2f_0}\delta H_{123,2}+
\left (2-{M\over r}\right )\delta H_{123}\right ]=0, \cr}
\eqn\diciannoven
$$
$$
\eqalign{
&\left (\htdz \right )_{,0}=0, \cr}
\eqn\diciannoveo
$$
$$
\eqalign{
&\left (\htdz \right )_{,3}=0, \cr}
\eqn\diciannovep
$$
$$
\eqalign{
&\left (\htdz \right )_{,2}+{1\over r}\left (2-{Q^2\over Mr}\right )e^{-2f_1}
\left (\htdz \right )=0, \cr}
\eqn\diciannoveq
$$
$$
\eqalign{
&\delta H_{123,0}+\delta H_{013,2}-\delta H_{012,3}=0, \cr}
\eqn\diciannover
$$
$$
\Delta=0.
\eqn\diciannoves
$$
Following the ansatz \sedicibis~on the perturbations of the metric we
shall divide our equations \diciannovea-\diciannoves~into two sets which
we will call
axial and polar. The equations for the axial (polar) perturbations will contain
only $\delta H_{1\alpha \beta}, \chi_0, \chi_2, \chi_3$ ($\delta H_{320}, \fa ,
\fb , \fc , \fd , \fx$).
The behaviour of the perturbations which is consistent with our previous
ansatz is:
$$
\cases{
\delta f(r,t,y) =\delta \tilde f(r)\cdot e^{i\omega t}\cdot
e^{ipy}, \cr
\delta \chi (r,t,y) =\delta \tilde \chi (r)\cdot e^{i\omega t}\cdot
e^{ipy}. \cr}
\eqn\ventidue
$$

\subsection{Axial Perturbations}

We now consider the equations for the axial perturbations only. From
\diciannoveq:
$$
\tilde \chi_3=  {Q^3\over M^2}e^{-2f_1}.
\eqn\ventitre
$$

Deriving \diciannovea, \diciannoveb~with respect to $y, r$, using
$\chi_{30,02}=\omega^2\chi_{23}+\chi_{20,03}$ and summing, we finally obtain:
$$
\eqalign{
{8\over k}\left [\left (e^{4f_1+2f_0}r\tilde\chi_{23}\right )_{,2}
e^{-2f_1+2f_0}r +re^{4f_0+2f_1}\tilde\chi_{23}+{Q^5\over M^2r^2}
e^{-2f_1+2f_0}\right ]_{,2} \cr
=-\omega^2\tilde\chi_{23}+p^2e^{2f_0}\tilde\chi_{23}.\cr}
\eqn\ventisei
$$
Defining:
$$
\tilde\chi_{23}=r^{-1}e^{-4f_1-2f_0}e^{-{1\over 2}\int dr^*~ X_1(r^*)}Y,
\eqn\ventisette
$$
where $X_1\dot = \sqrt{{8\over k}}e^{2f_0+f_1}-3\dr f_1$. Introducing
the ``tortoise" coordinate ${dr*\over dr}=\sqrt{{k\over 8}}{1\over r}e^{-2f_0
-f_1}$, we get:
$$
(\ddr +\omega^2)Y=V(r,Q,M)Y +J(r,Q,M),
\eqn\trentadue
$$
where:
$$
V=\left [{2\over k} \left (1-{3M\over r}+{3Q^2\over Mr}-{Q^4\over 4M^2
r^2}-{2Q^2\over r^2}+{5Q^4\over 4Mr^3}\right )e^{-2f_1}+p^2\right ]e^{2f_0},
\eqn\trentatre
$$
and:
$$
J={8Q^5\over kM^2r^{3/2}}\left (2-{3M\over r}-{Q^2\over Mr}+{2Q^2\over r^2}
\right  )e^{2f_0-3/8 f_1}.
\eqn\trentaquattro
$$

The effective potential \trentatre~is positive definite only for $Q=M$ and
it is plotted in \FIG\grafi{The effective potential in the extremal case
for $p^2=0$.} Fig.~\grafi.
In this limit:
$$
r^*=\sqrt{{k\over 2}}\left [\ln \left [\left ({r\over M}\right )^{1/2}\left
(1+e^{f_0}\right )\right ]-e^{-f_0}\right ],
\eqn\trentasei
$$
$$
\cases{r^*\rightarrow -\sqrt{{k\over 2}}e^{-f_0}\rightarrow -\infty&
as $r\rightarrow M^+$ \cr
r^*\rightarrow \sqrt{{k\over 8}}\ln \left ({4r\over M}\right )\rightarrow
+\infty&as $r\rightarrow +\infty$. \cr}
\eqn\trentasette
$$

In the extremal case the differential equation \trentadue~has an essential
singularity at $r=M$ and a regular singularity at $+\infty$.
$Y$ could be singular only in this two points.
By studying the dominant behaviour of the solutions around the two
singularities, we find:
\item\bullet
$r^*=+\infty$:
$Y=c_1e^{-{i\gamma \over \sqrt{2k}}r^*}+c_2e^{+{i\gamma \over
\sqrt{2k}}r^*}$, $\gamma=\sqrt{4+2k(p^2-\omega^2)}$.
\item\bullet
$r^*=-\infty$:
$Y=c_1e^{+i\omega r^*}+c_2e^{-i\omega r^*}+d(r^*)^{-5/2}$

\REF\wald{R.M.Wald\journal J.Math.Phys.&20(79)1056.}
Even if the $J$ term destroys the superposition principle, the finiteness
of the solution with respect to time evolution can still be proved along
the lines of Ref.\chandra, \wald. Let us not impose any particular time
dependence
for $Y$. Then for the resulting differential equation a conservation law is
easily found:
$$
\int ~dr^*~\left [\biggl \vert {\partial Y\over \partial t}\biggr \vert ^2
+\biggl \vert {\partial Y\over \partial r^*}\biggr \vert ^2 +V\bigl \vert
Y\bigr \vert^2\right ]=C-2\int ~dr^*~J{\cal R}e (Y).
\eqn\trentasettebis
$$
$C$ is a positive constant given the behaviour of $J$ which goes to zero for
$r_*\mapsto \pm\infty$. Moreover the $Y$ function is everywhere well behaved
as it follows from Fuch's theorems on the solution of this type of differential
equation and from the asymptotic analysis we have done earlier. The behaviour
of $Y$ around $r_*=\pm\infty$ is the same as that of the wavefunction coming
from the homogeneous Schr\"oedinger equation. From these condiderations it
follows that $\vert {\partial Y\over \partial t}\vert ^2$ is bounded.

\subsection{Polar Perturbations}

We now turn to the polar perturbations. Their equations are:
$$
\eqalign{&\tfa =\left ({3M\over 2r}-1\right )\left (1+{M\over 2r}\right )^{
-1}\tfc,  \cr}
\eqn\venti
$$
$$
\eqalign{
&\dr (\tfx ) =-\sqrt{{8\over k}}{M\over r}e^{f_0}(\tfa -\tfc), \cr}
\eqn\ventia
$$
$$
\eqalign{
&\dr (\tfa -\tfc ) =-{1\over 2}\sqrt{{8\over k}}{M\over r}e^{f_0}
\left [\tfa -\left (1+{2r\over M}\right )\tfc \right ], \cr}
\eqn\ventib
$$
$$
\eqalign{
&(\ddr +\omega^2)(\tfx )+\sqrt{{8\over k}}\left (1-{M\over 2r}\right )
e^{f_0}\dr (\tfx )=\biggl [p^2 \tfx + \cr
&+{8M\over kr}\left [\left (1+{M\over r}\right )\tfa +e^{2f_0}\tfc
\right ]\biggr]e^{2f_0}, \cr}
\eqn\ventic
$$
$$
\eqalign{
&(\ddr +\omega^2)(\tfa )+\sqrt{{8\over k}}\left (1-{M\over 2r}\right )
e^{f_0}\dr (\tfa )=\biggl [p^2(\tfx +\tfa )+ \cr
&+{8M\over kr}\left (1+{M\over r}\right )\tfa \biggr ]e^{2f_0},\cr}
\eqn\ventid
$$
$$
\eqalign{
&(\ddr +\omega^2)(\tfa -\tfx )+\sqrt{{8\over k}}\left
(1-{M\over 2r}\right )e^{f_0}\dr (\tfa -\tfx )= \cr
&=p^2e^{2f_0}(\tfx +\tfa ), \cr}
\eqn\ventie
$$
$$
\eqalign{
&(\ddr +\omega^2)(\tfc )+\sqrt{{8\over k}}\left (1-{M\over 2r}\right )
e^{f_0}\dr (\tfc )=\biggl [\biggl [p^2+{4M\over kr}\biggl (4+ \cr
&-{3M\over r}\biggr )\biggr ] \tfc
 +{4M^2\over kr^2} \tfa \biggr ]e^{2f_0}. \cr}
\eqn\ventif
$$
It is easy to see that the only consistent solution of such a system of
equations is $\delta\tilde f_i=\tfx=0, i=0,\cdots,3$. This is likely pointing
out that our system has some extra simmetries we have not been able
to isolate.

\subsection{Test Field}

The last computation we perform before concluding is that of the effective
potential of a test spinless boson in the background of our black hole
geometry. We thus study the equation:
$$
\nabla^2 \phi ={1\over \sqrt{-g}}\partial_{\mu}(\sqrt{-g}g^{\mu \nu}\partial_
{\nu}\phi)=0.
\eqn\cinotto
$$
This equation is separable and leads to:
$$
\phi (r,t,y,x)=\psi (r)X(x)Y(y)T(t)=\psi (r)e^{ik_1x}e^{ik_2y}e^{ik_3t}.
\eqn\cinnove
$$
In the extremal case we get:
$$
\partial^2_r \psi +\left [{2\over r-M}-{1\over r}\right ]\partial_r \psi
-\left [{k_2^2\over (r-M)^2}-{r(k_3^2-k_1^2)\over (r-M)^3}\right ]\psi =0.
\eqn\sessanta
$$
Using the ``tortoise" coordinate and the integrating factor $\psi =e^{-1/8f_0}
Z$, we finally get the Shr\"oedinger-like equation $(\ddr + k_3^2)Z=\widehat
VZ$ with
effective potential:
$$
\widehat V={M\over kr^3}\left [-{7M^2\over 2}+{11Mr\over 2}-(2+kk_2^2)r^2
\right ]+k_1^2+k_2^2.
\eqn\sesdue
$$
This effective potential has not a definite sign. When it is negative we have
the standard super-radiant behaviour and, using the notations of
Ref.\preskill, a finite mass gap for $k_2=0$ and $k_3^2< k_1^2$.
\refout
\figout
\end